\documentstyle{article}
\begin{document}

\begin{center}
{\LARGE Estimation of embedding dimension by}
\end{center}

\begin{center}
{\LARGE experimental data}
\end{center}

\begin{center}
{\Large Part 1}
\end{center}

\bigskip

\begin{center}
{Konstantin E. Bobrov,$^1$ Nickolas B. Volkov,$^1$\\
Alexander M. Iskoldsky, \footnote{Laboratory of modeling of
electrophysical processes, Institute of Electrophysics, Ural
Division of Russian Academy of Sciences, 620219, 34,
Komsomolskaya St., Ekaterinburg, Russia, E-mail:
ami@ami.e-burg.su}
Vyacheslav N. Skokov \footnote{
Laboratory of phase transitions and nonequilibrium processes,
Institute of Heatphysics, Ural Division of Russian Academy of
Sciences, 620219, 91, Pervomayskaya St., Ekaterinburg, Russia}}
\end{center}

\begin{abstract}
In connection with the investigations of initial stages of
appearance of turbulence in the current-carrying mediums and
also the investigations of relaxation oscillations in thin-film
bridges of high-temperature superconductor $Y Ba_2 Cu_3 O_{7-x}$
some problems of estimation of asymptotical parameters of
chaotic dynamic systems are considered. The mechanism of
measuring modes matching (discretization by time and by
amplitude at the limited buffer memory of registering device) to
the fixed procedure of signal handling is discussed. It is shown
that the problem of estimation of embedding dimension can be
reduced to the investigation of sufficiently simple phase
transitions at the integer-valued finite-dimensional lattices.
\end{abstract}

\section{Introduction}

As number of experimental works devoted to nonequilibrium phase
transitions is increasing, the problem of obtaining of
asymptotical estimations of natural chaotic systems parameters
becomes more urgent.

As a rule, records of long time series are considered to solve
this problem (see review \cite{Lille-94}).

This work was accomplished in connection with \cite{VolkIsk-90},
\cite{VolkIsk-93} where an initial stage of appearance of
turbulence at the electrical explosion of conductors is
investigated, and also in conection with \cite{Skokov-93-D},
\cite{Skokov-93-C} devoted to experimental investigations of
relaxation oscillations of current in the thin film of
high-temperature superconductive ceramic.

One of most important parameters of a dynamic system in the
steady-state mode is an embedding dimension --- number of
essential variables of the problem (dimension of a half-stream).
This value is of special interest for the theory too, in spite
of that for many nonlinear problems of mathematical physics
(including boundary problems of magnetohydrodynamic, see review
\cite{Chuesh-93}). As always, the problem of choice of number of
nonequilibrium modes that should be kept at the transition from
the base system of differential equations in partial derivatives
to the final system of ordinary differential equations remains
important. \footnote{In \cite{VolkIsk-90}, \cite{VolkIsk-93} the
three-mode approximation is used and the theoretical and the
experimental arguments in favour of this low-mode model are
presented.}

Almost all practical handling algorithms are based in one way or
another on the classical result \cite{Takens-81}, according to
which the necessary estimation can be made having the signal
record from only one of stream variables.

We also will use this result understanding that the Takens
theorem is valid only for a smooth mappings, but in the
experiment we deal with the signal discretizated by time and
by amplitude (by the space) as well.

It is well known (see for example review \cite{Blank-89}) that
the small perturbations of unstable dynamic systems can cause
qualitative change of their behaviour. For example, numerical
modeling isn't able to answer some questions because of rounding
errors. As to unstable cycles, the amplitude discretization can
lead to their stabilization and the time discretization --- to
their destruction.

Besides, we will actively use the construction that (as we
know) is first used by Gregory E. Falkovich in the article
\cite{Falk-85} (see also \cite{Falk-89}), devoted to
experimental estimations of embedding dimension and fractal
dimension of chaotic attractors, appearing at the initial stage
of hydrodynamic turbulence in the Couette flow.

We will dwell on this construction in more details noting only,
that there is used the almost obviously fact: if the next after
the marked (reference) zone of signal single reading
is functionaly connected to the previous reading \footnote{It take
place if the number of readings in the reference zone is equal
or greater than the number of independent variables of dynamical
system}, than meeting the similar zone of signal we can hope that
reading immediately following the referenced one is similar to
single one.

In the first part of this work the description of experiment is
presented and the methodical questions concerning to the
estimation of embedding dimension by experimental data are
discussed: accounting of finite accuracy of signal amplitude
measurements makes impossible the direct application of known
Takens algorithm.

\section{Experiment}

The physical results obtained in experiment presented
partly in \cite{Skokov-93-D}, \cite{Skokov-93-C}. We will
discuss only methodological side of it.  The relaxation
oscillations of current and voltage in the  thin
film high-temperature superconductor bridges $Y Ba_2 Cu_3 O_{7-x}$
(critical temperatures --- $T_c = 86 - 88$ K, density of
critical current --- $j_c = 10^5 - 10^6$ A/cm$^2$ at 77 K and
absence of external magnetic field) were investigated. The
samples were fabricated by direct current magnetron sputtering
with target of stehiometric consistant onto the substrates made
of monocrystals $Sr Ti O_3$ É $Zr O_2 (Y_2 O_3)$. The dimensions of
bridges were: thickness --- $\sim$ 0.3 $\mu$m, width --- 0.7 -
1.0 mm, length --- 1 - 7 mm.

\begin{figure}[h]
\caption{Schematic diagram of measurements. At the obtaining of
steady-state voltage-current characteristic $L_0 = 0$.}
\end{figure}

The sample was connected to the electrical circuit according to
the schematic diagram shown in the Fig. 1 and was plunged into the
dewar with the liquid nitrogen. The measurements of current and
voltage was made by using standard four-points scheme. The
oscilloscope C9-8 was used for registration of oscillogramms. It
has 8-digits analog-digital converters and the buffer size for
signals recording is 2048 bytes. The steady-state
voltage-current characteristic shown in Fig. 2 were obtained in
the mode with fixed source voltage and the connecting the sample
into the circuit in series with load resistor (without
inductance). The portion of voltage-current characterictic wit
negative differential resistance shown in the Fig. 2 defines
development of instability \cite{Skokov-93-C}. The oscillogram
of voltage in circuit with inductance ($L_0 \not= 0$) is shown
in Fig. 3. It is obvious that we deal with well expressed
periodic mode.

\begin{figure}
\caption{a) The steady-state voltage-current characterictic;
b) The dynamic voltage-current characteristic in periodic mode
($L_0 \not= 0$).}
\end{figure}

Note that current $I(t)$ is directly among of the set of
phase variables governing the system dynamic. It comes
in the form of the integral of $j_\rho$ --- laminar consistant
of current taken over cross-section of superconductive
film (see \cite{VolkIsk-90}, \cite{VolkIsk-93}):

$$I(t) = \int \limits_{S}^{} j_\rho ds$$

The voltage $U(t) = \int \limits_{l}^{} E_l dx$, where $E_l$ is
the longitudinal component of electrical field on the film's
surface, $l$ is the distance between the measuring electrodes.

In the above considerations $I(t)$, $U(t)$ are continuous and
rather smooth functions of time.

The analog signals $I(t)$, $U(t)$ are applied to the input
of measuring device that makes encoding (discretization by
amplitude) of both signals with the regular time step
$\tau$. On completing of one measuring cycle we have
the ordered array of code pairs, filling in the buffer
memory of oscilloscope. Every code can be interpreted as integer
positive number in the range $[0, 2^m - 1], m=8$. It is
possible to assume that this number represents the result of
integer division of "instant" value of corresponding measuring
signal by the interval between the levels of it's discretization.

\begin{figure}
\caption{The oscillogram of voltage in HTSC film. The selected
fragment shown in Fig. 6.}
\end{figure}

\begin{figure}
\caption{The normalized dynamic voltage-current characteristic
in the periodic mode. The selected fragment is shown in Fig. 5.}
\end{figure}

\begin{figure}
\caption{The fragment of Fig. 4.}
\end{figure}

Almost all possible range of codes is usually in effect. At
these conditions the aprioral characteristic of accuracy of
measurement of signals by amplitude (by space) is the value
$\epsilon = 2^{-m}$ \footnote{The situation like this (so called
{\it regular} $\epsilon$ --- {\it discretization}) appears
due to the errors of rounding at the numerical modeling of
dynamic systems \cite{Blank-89}}. At $\epsilon \rightarrow 0$ we
can assume that the process of measurement reduces to
time discretization of continuous signals.

The so called dynamic voltage-current characteristic of film
is shown in Fig. 4. Every it's point corresponds to the pair of
codes $(I, U)$ obtained at the same moment. This figure can be
considered as a point mapping of a segment of phase trajectory
of the system onto the plane of measurements. The large-scaled
fragment of Fig. 4 is shown in the Fig. 5. We can see that the
points of trajectory of discretizated system are located at the
nodes of the integer-number mesh (lattice).

One more effect of space discretization is the appearing of the
consecutive identical codes in sequences of codes (flat
portions in oscillograms) (see Fig. 6).

What hinders the situation is that the perturbations effected by
the measurement procedure are not lone in the system.

{\it On amplitude measurements.} The amplitude of signal and
the average point of code scale (the value of code at the zero
signal) we select so that the realization will not contain codes
located at the border of the scale. On the choice made the
system trajectory calls upon only the internal points of
coding space. If $m \geq 8$ this choice does not lead to
significant loss of accuracy. The corresponding component of
error is as a rule in the range

$$ \epsilon = \left( \frac{(2^{m-1} - 1)^{-1}}{2},
\frac{(2^{m-2} - 1)^{-1}}{4} \right).$$

{\it On choice $\tau$ --- time step value.} The limitation
of buffer memory size essentially influences this choice.

\begin{figure}
\caption{Fragment of the oscillogram of voltage from Fig. 3.}
\end{figure}

Too fine quantization will cause the uneffective filling of $N$
volume by the long sequences of identical repeating codes in
limiting case by single code. To make $\tau$ greater than
characteric correlative time is uneffective too.

In actual practice, at least at the initial stage of
investigation the principle of "peer interest to space and to
time" is in effect and the step magnitude $\tau_0$ is set to
make the measurement error of characteric time interval (in this
case --- of period $T$) to be close to $\epsilon$. For example,
if $\epsilon = 2^{-8}$, $N=2^{11}$ bytes than on setting $\tau$
in accordance with aforementioned principle it will be
registered $K_0 \sim N \epsilon$, $K_0 \sim 8$ periods of
oscillations (See Fig. 3).

To check if it is possible to consider the trajectory as a cycle
it is necessary to have $K_{min} \geq 2$, i. e. $\tau_{min} \geq
0,25 \tau_0$. One can evaluate $K_{max}$ specifying the limit
value of error in the estimation of period length. So, for
$\epsilon_{max} = 3$\%, we have $K_{max} \leq 32$, i. e.
$\tau_{max} \leq 4 \tau_0$.

\section{Handling of data}

At the handling of experimental data we will try not to go off
the frames of integer numbers set that is inherent to this data.
In some cases it makes possible to reduce the problem to
investigation of rather simple structural phase transitions in
the integer-number finite-dimensional lattices. The benefits of
this method, we think, make up for possible inconveniences. In
connection to this circumstance we will consider the main
procedures in some simple model examples.

\subsection{Integer-number model}

\begin{figure}
\caption{The model signal $d=3$, $n=8$.}
\end{figure}

Let us assume the model signal fragment of which is shown in
Fig.  7:

\begin{equation}
U(t) = \left\{ 01212321012123... \right\}
\end{equation}

$$U(t) = \left\{U_j\right\}, \quad j = 1, ..., 14, ..., N.$$

The first step envisages the building from (1) vectors (corteges)
with the length $k$ by the next algorithm:

i) Fixing initial vector

\begin{equation}
\xi^k_1 = \left\{ \xi^k_{11} = U_1, ..., \xi_{1k} = U_k \right\}.
\end{equation}

In our example $\xi^1_1 = \left\{ 0 \right\},
\xi^2_1 = \{ 01\},
\xi^3_1 = \left\{ 012 \right\}$.

ii) Using mapping according to which by known $\xi^k_j$
is builded $\xi^k_{j+1}$

\begin{equation}
f^j : \left\{ \xi^k_{(j+1),1} = \xi^k_{j,2} \right\}, ...,
\left\{ \xi^k_{(j+1),i} = \xi^k_{j,(i+1)} \right\}, ...,
\left\{ \xi^k_{(j+1),k} = U_{j+k+1} \right\},
\end{equation}

$$j = 1, ..., N-k.$$

As a result, the sequence appears with length of $(N-k)$
$k$-components vectors. In our example for $k=1$ it is
just starting sequence, for $k=2$ --- 13 pairs:

\begin{equation}
< (01)(12)(21)(12)(23)(32)(21)(10)(01)(12)(21)(12)(23) >,
\end{equation}

For $k=3$ --- 12 triplets.

\begin{equation}
< (012)(121)(212)(123)(232)(321)(210)(101)(012)(121)(212)(123) >
\end{equation}

It is necessary to find $k=d$ so that the cortege $< \xi^d_1,
..., \xi^d_n >$ is a cycle of period $n$ with the following
properties:

a) there are no identical vectors in the limits of a period

\begin{equation}
\xi^d_i \not= \xi^d_j , \qquad i \not= j , \quad i, j < n;
\end{equation}

b) the vectors separated one from another with period exactly
are identical ones

\begin{equation}
\xi^d_{l+nK} = \xi^d_l,
\end{equation}

here $K$ is the ordinal number of a period.

As it was specified, the sequence (4) is not a cycle because it
contains repeating pairs. Nevertheless, (5) is a cycle of
period $n = 8$.

The desired embedding dimension is $U(t)$ $d = 3$. Note, that
here $d$ is integer-valued. It should be considered as a right
border of noninteger-valued dimensions that can characterize
the dynamics of system in the chaotic modes of its functioning.

It is clear that in our case one has to understand the identity
(nonidentity) in a certain sense. Before to do it, let's recall
that $(i+k+1)$th reading of main sequence $U_{i+k+1}$ directly
follows $\xi^k_i$, and $U_{j+k+1}$ follows $\xi^k_j$.

Let us settle some criterion of vectors $\xi^k_i$, $\xi^k_j$:
$dist(\xi^k_i , \xi^k_j)$ resemblance and assume the small
$\epsilon$, generally speaking, depending on $i, j$. The
criterion that on $k \geq d \quad$ $(k+1)$-th reading is a
function of $k$ previous readings can be formalized in the
following way.

If on $k \geq d$ there exists an $\epsilon$-vicinity of zero
$(\epsilon_\rho, \epsilon_r)$ so that from

\begin{equation}
dist(\xi^k_i , \xi^k_j) \leq \epsilon_\rho (i, j)
\end{equation}

follows

\begin{equation}
dist(U_{i+k+1} , U_{j+k+1}) \leq \epsilon_r (i, j),
\end{equation}

then $k=d$ is an desired evaluation.

In cited above \cite{Falk-85} Euclidean distances are set as a
degree of proximity

\begin{equation}
\rho^k = dist(\xi) = \mid \mid \xi^k_i - \xi^k_j \mid \mid ,
\end{equation}

and

\begin{equation}
r^k = dist(U) = \mid \mid U_{i+k+1} - U_{j+k+1} \mid \mid ,
\end{equation}

where $\mid \mid \bullet \mid \mid$ is the norm of corresponding
differences. It is not very convenient because $\rho^k$, $r^k$
appear to be real numbers albeit initial data belongs to the
field of integer numbers. According to the assertion at the
beginning of section 3, we will use a module (not a norm) of
corresponding differences as a degree of proximity (unless
specified otherwise):

\begin{equation}
\rho^k = \mid \xi^k_i - \xi^k_j \mid,
\end{equation}

and

\begin{equation}
r^k = \mid U_{i+k+1} - U_{j+k+1} \mid .
\end{equation}

Using the algorithm (1), the algorithm i)-ii), and also
(12)-(13) we build the plots $r^k (\rho^k)$ consecutively
comparing the selected pair $<\xi^k_l, U_{l+k+1}>$ to the
rest of pairs $j \geq l$.

This procedure makes a mapping of the initial trajectory (1) at
$k=1$ into the square $\mid 3 \times 3 \mid$, at $k=2$ ---
into rectangle $\mid 3 \times 6 \mid$, at $k=3$ ---
into $\mid 3 \times 9 \mid$, and so on. At the first step $j=l$,
so that the trajectory $(\rho, r)$ starts from the point $(0, 0)$
and on completing the cycle comes back to this point.

Varying $l=1, ..., n$ for each $k=1, 2, 3$ we get $n$ meshes
showed in Fig. 8 - 17. From the point of view of topology all
meshes represent multy-beam stars with centre in the point with
coordinate $(k, 1)$.

\begin{figure}
\caption{Meshes for the signal in Fig. 7; $k=1$, $l=1 \div 4$.}
\end{figure}

\begin{figure}
\caption{Meshes for the signal in Fig. 7; $k=1$, $l=5 \div 8$.}
\end{figure}

\begin{figure}
\caption{Meshes for the signal in Fig. 7; $k=2$, $l=1 \div 2$.}
\end{figure}

\begin{figure}
\caption{Meshes for the signal in Fig. 7; $k=2$, $l=3 \div 4$.}
\end{figure}

\begin{figure}
\caption{Meshes for the signal in Fig. 7; $k=2$, $l=5 \div 6$.}
\end{figure}

\begin{figure}
\caption{Meshes for the signal in Fig. 7; $k=2$, $l=7 \div 8$.}
\end{figure}

\begin{figure}
\caption{Meshes for signal in Fig. 7; $k=3$, $l=1 \div 2$.}
\end{figure}

\begin{figure}
\caption{Meshes for signal in Fig. 7; $k=3$, $l=3 \div 4$.}
\end{figure}

\begin{figure}
\caption{Meshes for signal in Fig. 7; $k=3$, $l=5 \div 6$.}
\end{figure}

\begin{figure}
\caption{Meshes for signal in Fig. 7; $k=3$, $l=7 \div 8$.}
\end{figure}

Let choose as a

\begin{equation}
\epsilon_\rho = min \left\{ \rho_0, \rho_S \right\},
\end{equation}

where $\rho_0 = r_0$ is abscissa of the centre of the star at
$k=1$, and $\rho_S$ evaluate as a point of intersection of the
line $r=\rho$, drawn from the point of contact of the plot with
the axis of ordinates to the point of intersection with its
top delimiting border (see Fig. 8). For this model $\rho_S = r_0
= 1$.

Beginning from the Fig. 8 - 9 $(k = 1)$ in two cases $(l=1, 6)$
of eight the structure of mesh meet the demands of
criterion (8), (9). Namely, in the area $\epsilon_\rho \leq 1$
all its points do not fall beyond the border line $r=\rho$.

In Fig. 10 - 13 $(k=2)$ four meshes from eight possess this
property, in Fig. 14 - 17 $(k=3)$ all of eight. As to the rest
of meshes in Fig. 8 - 9, 10 - 13, presence of the "bad" node
with coordinate $(0, 2)$ within their structure prevents the
meeting of criterion (8), (9).

Thus the {\it structural phase transition in a discrete
finite-dimensional lattice} is present. The subset of meshes
(number of meshes $n_a$) whose structure comprises the nodes
located beyond the line $r=\rho$ acts as initial stage.
$n_\beta$ meshes represents the new $\beta$-stage and in the sum
$n_\alpha + n_\beta = n$ where $n$ is a period of a cycle. For
this phase transition $k$ act as the parameter of order and
$k=d$ is its critical value. Note two more peculiarities:  a)
the distance  between the location of "bad" node (0, 2) in the
initial stage and "good" node (2, 2) in the new stage is
constant along the lattice (gap) and equal 2; b) node with
coordinate (0, 1) is always empty.

The model signal is an ideal in many respects. For example, if
we take (1) and trace the dynamics of volume of elementary
cell of phase space

\begin{equation}
\Delta V (l) = \prod_{i=1}^{3} \mid \left \{ \xi^3_{l+1, i} -
\xi^3_{l, i} \right \} \mid ,
\end{equation}

$l = 1, ..., n, ... N-3$, we will come to the conclusion that
$\Delta V (l)$ is a unit cube propagating without deformation.

There is no additive noise in (1) and the amplitude transitions
occur only between two contiguous levels. As a result in all of
plots the centre of the star is located at the ordinate closest
to the zero $r_0 = 1$. In general, $r_0 > 1$.

There never appear two (and more) identical sequential codes in
the model example (1) but it is by no means nearly so in
practice. For example, Fig. 3 contains (see Fig. 6) long (up to
32 elements) sequences of identical codes --- this circumstance
can radically deteriorate the quantity of estimation $d$
according to this algorithm.

As illustration to the last assertion we consider the sequence:

\begin{equation}
\left \{ 011221122332211011 ... \right \},
\end{equation}

obtained from (1) by repeating twice of all codes besides 0.
The meshes for (16) are shown in Fig. 18. The following changes
occured as against Fig. 8 - 17:

--- it is no longer a star with well emphasized centre;

--- the loops corresponding to the consecutive identical pairs
$(\rho , r)$ appeared into the graph structure besides simple
cycles and short sequences;

--- the trajectory repeatedly visits the axis of ordinates at
the point (0, 1) that did not occure before. Nevertheless, the
phase transition takes place at $k=3$.

\begin{figure}
\caption{Meshes for sequence (16).}
\end{figure}

\begin{figure}
\caption{Meshes for signal in which as opposed to (16) the third
unity is added to the sequence of unities following the first
zero.}
\end{figure}

\begin{figure}
\caption{The illustration of systematic error stipulated by
finiteness of realization length $N$.}
\end{figure}

The sitution changes drastically when the maximum length of a
sequence of identical codes in signal $\Delta l_{max} \geq d$.
From this moment the $\Delta l_{max}$ governs the estimation of
$d$.  Fig. 19 presents the mesh $(k=3, l=2)$ for the signal that
differs from (16) by adding of third unity  into the sequence of
unities following the zero. The mesh in Fig. 19 is shifted by
one step to the left as against Fig. 15. This has the effect of
shifting the good node $(2, 2)$ to the point with coordinates
$(1, 2)$ converting it to the bad one.

Note some effects deteriorating the quantity of estimation of $d$
due to increasing {\it systematic} part of error. With the
increasing of $k$ the space taken by the plot is also increasing
(number of nodes of the coordinate mesh grows). On fixing: (1)
volume of realization and (2) size of zone close to zero $\rho =
[0, \epsilon_\rho]$ that we are interested in, the density of
depicting points decreases.

As illustration we consider this effect on example of sequence
$N = 2048$ bytes generated by the pseudo-random number
generator for $k = 1, 3, 6$ (Fig. 20). While at $k = 1$
depicting points evenly cover the space of the plot, on
increasing of $k$ they contract into the narrow band in its
centre. \footnote{One can "lose" rarefied zone near
the zero starting calculation $r(\rho)$ from $j=l$ (not
from $j=l+1$). In this case (using standard procedures of
of plot drawing), the absciss of the left low corner will be
defined by minimum $\rho \not= 0$ in the particular
realization.}.

Completing the subsection (3.1)  we address again to
\cite{Falk-85} (see Introduction) where the concept of
uppermost envelope for the plots $(\rho, r)$ is adopted and
its monotonicity near the zero is stated. Actually, in the limit
$\epsilon \rightarrow 0$ one may use term envelope $r(\rho)$. At
$k \geq d$ in the zone of a small $\rho < \epsilon_\rho$ scales
from

\begin{equation}
\rho^d = \mid \mid \xi^d_i - \xi^d_j \mid \mid \rightarrow 0,
\end{equation}

follows

\begin{equation}
r^d = \mid \mid U(t_{i+d+1}) - U(t_{j+d+1}) \mid \mid
\rightarrow 0.
\end{equation}

In real life $\epsilon \not= 0$; the sequences of repeating
codes may be present in signal and the longest of them directs
eventually the quantity of estimation of embedding dimension by
Takens-Falkovich algorithm.

\section{Conclusions}

Thus, we showed that signal amplitude discretization makes it
impossible the direct application of classical Takens-Falkovich
algorithm for the direct estimation of embedding dimension. One
can note that various signal sections are disparate in the sense
that they are able to yield the different estimations for $d$.
This makes us to choose the fixed (depending on signal
properties, accuracy of its amplitude and realization length
measuring) time discretization step and to solve the phase
transition problem onto the lattice (in accordance with the
algorithm presented in the section 3).

In the second part of this paper we will consider some
techniques of obtaining the desired estimation by the example of
Lorenz system (operating in various modes) and we are going
to present the results of processing of experiment described in
section 2 obtained by means of program taught by Lorenz system.

\end{document}